\documentclass[12pt]{iopart}

\usepackage{graphicx}

\makeatletter
\def\lsim{\mathrel{\mathpalette\gl@align<}}
\def\gsim{\mathrel{\mathpalette\gl@align>}}
\def\gl@align#1#2{\lower.6ex\vbox{\baselineskip\z@skip\lineskip\z@
    \ialign{$\m@th#1\hfil##\hfil$\crcr#2\crcr\sim\crcr}}}
\makeatother

\begin{document}

\title[Phonon-assisted tunneling in interacting suspended
 SWNTs]{Phonon-assisted tunneling in interacting suspended single wall
 carbon nanotubes}

\author{Wataru Izumida$^{1,2}$ and Milena Grifoni$^{3}$}

\address{$^{1}$Department of Physics, Tohoku University, Sendai
980-8578, Japan \\
$^{2}$Kavli Institute of Nanoscience, Delft University of Technology,
Lorentzweg 1, 2628 CJ Delft, The Netherlands\\
$^{3}$Department of Physics, University of Regensburg, D-93040
Regensburg, Germany}

\ead{izumida@cmpt.phys.tohoku.ac.jp}

\begin{abstract}
Transport in suspended metallic single wall carbon nanotubes in the
presence of strong electron-electron interaction is investigated. We
consider a tube of finite length and discuss the effects of the coupling
of the electrons to the deformation potential associated to the acoustic
stretching and breathing modes. Treating the interacting electrons
within the framework of the Luttinger liquid model, the low-energy
spectrum of the coupled electron-phonon system is evaluated. The
discreteness of the spectrum is reflected in the differential
conductance which, as a function of the applied bias voltage, exhibits
three distinct families of peaks. The height of the phonon-assisted
peaks is very sensitive to the parameters. The phonon peaks are best
observed when the system is close to the Wentzel-Bardeen singularity.
\end{abstract}

\maketitle

\section{Introduction}

Carbon nanotubes are new prototypical candidates for one-dimensional
nanoconductors.  As such, electron transport through such macromolecules
has been investigated extensively.  Single wall carbon nanotubes (SWNTs)
behave as metallic ($n-m=3l$, $l$ is an integer) or semiconducting
conductors ($n-m \ne 3l$) depending on their wrapping vector,
$(n,m) \equiv n \overrightarrow{a_1} + m \overrightarrow{a_2}$, where
$\overrightarrow{a_1}$, $\overrightarrow{a_2}$ are the graphene
primitive lattice vectors.
\cite{rf:Saito98,rf:Wildoer9801,rf:Odom9801}. Moreover, Coulomb blockade
phenomena, the Kondo effect, and Tomonaga-Luttinger liquid properties
have been experimentally demonstrated
\cite{rf:Tans9704,rf:Bockrath9703,rf:Nygard0011,rf:Bockrath9902,rf:Negard9904}.
Such measurements confirm the relevance of electron correlations effects
in SWNTs at low energies as predicted in earlier theoretical studies
\cite{rf:Egger9712,rf:Kane9712,rf:Egger9806}.
Recently, suspended nanotubes have been fabricated, and the transport
properties have been measured by several groups
\cite{rf:Nygard0110,rf:Franklin0207,rf:Jarillo-Herrero0405,rf:Sazonova0409,rf:LeRoy0411}.
Since the suspended nanotube is free from the substrate, it is expected
that it can easily vibrate \cite{rf:LeRoy0411,rf:Babic0309}. At the same
time, some theoretical works have started to investigate the effects of
the mechanical motion on the electronic transport properties of
interacting SWNTs
\cite{rf:Komnik0209,rf:Sapmaz0307,rf:Martino0306,rf:Blanter0409}.  While
Ref. \cite{rf:Sapmaz0307,rf:Blanter0409} focus on Coulomb blockade
phenomena, the SWNT is treated as a Tomonaga-Luttinger liquid
\cite{rf:Haldane81} with periodic boundary conditions (valid for very
long nanotubes) in Ref. \cite{rf:Komnik0209,rf:Martino0306}.
However, typical SWNT lengths are of the order of sub-micron to
micro-meters, yielding a mean level spacing for the electronic energy
levels in the ${\rm meV}$ energy range. This discreteness of the energy
levels can be clearly seen in low temperature measurements, e.g. by
measuring the excitation spectrum of a SWNT quantum dot
\cite{rf:Sapmaz05}. In finite length SWNTs, the quantized nature of the
vibrational modes of a SWNT should be considered as well at low enough
energies.

In this paper, we investigate the effects of the mechanical vibrations
on the electronic properties of a doubly clamped SWNT of finite length
within a full quantum mechanical treatment of the low energy
excitations.
Even if the metallic condition $n-m=3l$ is satisfied, small energy gaps
$\sim 10 {\rm meV}$ appear due to curvature effects. The armchair
nanotubes ($n=m$), however, still stay metallic because of their high
symmetry \cite{rf:Ouyang0104}.  Therefore we consider the armchair SWNTs
in this paper.
The interacting electrons in the nanotubes are considered within the
standard bosonization framework of the Tomonaga-Luttinger liquid model
with open boundary conditions \cite{rf:Fabrizio9506,rf:Mattsson9712},
where the electronic excitations (charged and neutral plasmons) have
bosonic character.
The low energy vibrations of the nanotube can be described in terms of
acoustic phonons, associated to the longitudinal stretching mode and to
the transverse twisting and breathing modes \cite{rf:Suzuura0205}. The
coupling between electron and phonons is described in terms of the
deformation potential caused by the acoustic modes. In our model this
yields a bilinear bosonic coupling between phonons and plasmons.
In Ref. \cite{rf:Martino0306}, the acoustic modes have been
treated within a continuous elastic model, and have been
integrated out in order to obtain a phonon-mediated retarded
electron-electron interaction. It was shown that while the
twisting mode can be neglected at low energies, the effect of the
breathing mode and stretching modes should be considered. In
particular, retardation effects of the breathing mode can be
neglected, and an effective renormalization of the
electron-electron interaction strength is found with possibly
attractive interactions. In contrast, retardation effects due to
the stretching mode must be retained, so that, due to the combined
effects of stretching and breathing modes, the Wentzel-Bardeen
singularity, describing a superconducting instability of the
electron-phonon system \cite{rf:Loss9410}, can be reached. In this
paper, we consider the case in which electron {\em as well as}
phonon modes are quantized, and look to the energy spectrum of the
interacting electron-phonon system. As in Ref.
\cite{rf:Martino0306}, the major effect of the breathing mode is
to renormalize the electron-electron interaction strength.  On the
other hand the electronic as well as the stretching-phonons
parameters get renormalized by the mutual interaction, yielding
"electron"-like and "stretching-phonons"-like excitations. Such
electron-like excited levels appear as large peaks in the
differential conductance, while stretching-phonon-like assisted
processes appear as small peaks beside the large peaks. The phonon
peak height is very sensitive to the system parameters, and it
becomes larger and larger when the value of the charged plasmons'
velocity is close to the sound velocity, i.e. when the
Wentzel-Bardeen instability \cite{rf:Loss9410} is approached.
Thus, the experimental observation of the phonon peak would be an
indication of the strong fluctuation of the superconducting order
parameter, which corresponds to the Wentzel-Bardeen singularity,
in SWNTs.
At temperatures higher than the level spacing, the discreteness of the
peaks is washed out. The differential conductance then shows the
characteristic Tomonaga-Luttinger power-law dependence on voltage, with
exponents modified by the electron-phonon interaction.

\section{Low-energy Hamiltonian }

To start with, we consider a suspended, doubly clamped metallic SWNT
whose fixed ends are at $x=0$ and $x=L$.
We wish to develop a low energy theory for suspended metallic
SWNTs, and restrict ourselves to the two lowest linear in $k$
electronic subbands. Moreover, in experiments the nanotubes are
typically intrinsically $p$-doped such that the Fermi level is
shifted away from half-filling.
Long-range Coulomb interaction is dominant in isolated SWNTs,
because the electrons are spread out around the circumference of
the tube, and the probability of two electrons to be near to each
other is of the order $1/N$, where $N$ is the number of atoms on
the circumference. The interaction is usually screened on a length
determined either by the tube length or by the distance to nearby
gates. The Coulomb interaction causes Umklapp, backward and
forward scattering processes among the electrons. Away from
half-filling Umklapp scattering can be neglected. We also
disregard backscattering processes, which is a valid approximation
if the tube radius is not too small \cite{rf:Egger9806}. The
forward scattering processes can be fully included within a
Tomonaga-Luttinger (TL) model for SWNTs
\cite{rf:Egger9712,rf:Kane9712,rf:Egger9806}, yielding the TL
Hamiltonian
\begin{eqnarray}
 H_{\rm el} & = & H_{N} + \sum_{j} H_{j}~, \label{eq:H_e} \\
 H_{j} & = &
  \frac{v_{j}}{2} \int_{0}^{L} dx
  \left[ g_{j} \Pi_{j}^{2}(x)
   + \frac{1}{g_{j}} \left( \partial_{x} \phi_{j}(x) \right)^{2} \right],
  \label{eq:H_j}
\end{eqnarray}
where the index $j={\rm c+, s+, c-, s-}$ counts the four excitation
sectors for total charge, total spin and relative (with respect to the
two electronic subbands) charge and relative spin, respectively.
The first term $H_{N} = \sum_j E_{j} N_{j}^{2} / 2$ in
Eq. (\ref{eq:H_e}) represents the ground state energy. The quantities
$E_j$ are the
energies for total/relative charge and spin for a given number of
electrons $N_{\rm c \pm}$ and given spin $N_{\rm s \pm}$.
Specifically, $E_{\rm c+}$ contains the charging and single particle
energies and is taken as a free-parameter hereafter, in order to include
screening effects of nearby gates.  For the remaining modes is
$E_j=\varepsilon_{0}/4$, with $\varepsilon_0=\hbar v_{\rm F}\pi/L$ the
single particle level spacing. Here we denoted with $v_{\rm F}$ the
Fermi velocity, while $L$ is the SWNT length.
For SWNTs the term $\varepsilon_0 \delta N_{\rm c-}$ should be added in
$H_{N}$, where $\delta$ is a small off-set caused by the discrepancy
of the discretized energies between the two bands.
The gate potential term $- e N_{\rm c+} V_{\rm G}$ is also added in
the Hamiltonian.
The second term in Eq. (\ref{eq:H_e}) describes the collective bosonic
excitations of the one-dimensional interacting electron system.
$\Pi_{j}(x)$ and $\phi_{j}(x)$ play the role of the momentum and
displacement field operators, respectively, and they are conjugate
variables of each other, i.e., $[\phi_{j}(x),\Pi_{j'}(x^{'})]=i \hbar
\delta_{jj'} \delta(x-x^{'})$.
Due to the interaction, the excitations propagate with renormalized
velocity $v_{j}=v_{\rm F}/g_{j}$, where $g_{j}$ is the interaction
parameter for the $j$-sector.
Because the Coulomb interaction describes an interaction among total
electronic densities, only the total charge sector is affected while the
other sectors are neutral sectors, i.e., $g_{\rm c+}<1$ and $g_{j}=1$
($j={\rm s+, c-, s-}$), respectively. Hereafter we denote $g=g_{\rm c+}$
and we call $j={\rm c+}$ the charge sector simply.
The interaction parameter is roughly estimated as $g=1/\sqrt{1+U/\hbar
v_{\rm F} \pi}$ with $U=8e^{2} \log R_{\rm sc}/R$. Here $R_{\rm sc}$ is
a phenomenologically introduced screening scale, long compared with the
tube radius $R$. It can be estimated to be $R_{\rm sc}= {\rm min}\{L,d
\}$, where $L \simeq 1 {\rm \mu m}$ is a typical SWNT length,
and $d$ is the distance to a possibly present gate electrode.
The interaction parameter is estimated to be $g \simeq 0.2$ for $R_{\rm
sc} \simeq 1{\rm \mu m}$ \cite{rf:Kane9712}. The screening caused by a
gate electrode nearby the nanotube is typically more effective in
screening, which pushes $g$ closer to the non-interacting value, $g=1$.

We consider next the effects of the vibrations.
The relevant phonon modes at low energies are those for which the
quantized (due to the periodic boundary conditions in the circumference
direction) transverse momentum is zero.  There are three modes:
stretching, twisting and breathing modes.  In terms of the deformation
potential, the transverse-acoustic twisting mode does not contribute at
first order in the displacement \cite{rf:Martino0306}, so that it will
no longer be considered in the rest of this work.
The transverse acoustic breathing mode has a finite frequency
$\omega_{\rm B}$, of the order of $(0.14/\hbar R) {\rm meV}\AA$ in the
long wavelength limit. The energy scale is thus larger than that of the
low energy electrons in the vicinity of the Fermi energy. However, as we
shall discuss later, this mode strongly renormalizes the energy of the
charge sector.
In the following, we start by explicitly considering the effects of the
coupling to the longitudinal stretching mode. We start from a continuum
model for the vibrations \cite{rf:Suzuura0205}, such that the stretching
mode corresponds to the longitudinal acoustic mode of the
one-dimensional continuum. At a second stage of the calculation also the
effect of the breathing mode will be discussed.
The Hamiltonian for the longitudinal phonons and the electron-phonon
interaction are written as
\begin{eqnarray}
 H_{\rm L,ph} & = &
  \frac{1}{2}
  \int_{0}^{L} dx
  \left[ \frac{1}{\zeta} P^{2}(x)
   + \zeta v_{\rm st}^{2} \left( \partial_{x} u(x) \right)^{2} \right],
  \label{eq:H_ph} \\
 H_{\rm el-L,ph} & = &
 c
 \int_{0}^{L} dx
 \rho(x) \partial_{x} u(x),
 \label{eq:H_el-ph}
\end{eqnarray}
where $\zeta$ is the carbon mass per unit length, $v_{\rm st}$ is
the sound velocity of the stretching mode, $c$ is the
electron-phonon coupling constant, and $\rho(x)= 2 \partial_{x}
\phi_{{\rm c}+}(x) / \sqrt{\pi \hbar} + N_{\rm c+}/L$ is the
electron density.
$P(x)$ and $u(x)$ are the momentum and displacement field operators,
obeying the commutation relation $[u(x),P(x')]=i \hbar \delta(x-x')$.

Before we proceed, let us discuss the relation between the
coupling constant $c$ of our one-dimensional model and the
three-dimensional nature of the electron-phonon coupling in SWNTs.
For a SWNT the electron-phonon interaction is written as
\cite{rf:Suzuura0205}
\begin{eqnarray}
 H_{\rm el-ph} & = &
\int_{0}^{2 \pi R} dy \int_{0}^{L} dx \rho(x,y) V(x,y),
\end{eqnarray}
where the $y$ axis is along the circumference, $\rho(x,y)$ is the
density of electrons and  $V(x,y)$
is the deformation potential.
For the stretching phonon mode  $V(x,y)=2 c'
\partial_{x} u_x \mu / (B + \mu)$ has no $y$-dependence. Here $u_{x}$ is the displacement field,
and $B$,  $\mu$ and $c'$ are the bulk modulus,  shear modulus and
deformation potential for a graphene sheet, respectively.
Therefore, with $\rho(x)=\int_{0}^{2 \pi R} \rho(x,y) dy$, we have
the relation $c = 2 c' \mu /(B+\mu)$.

Since we want to investigate transport properties of the coupled
electron-phonon system in a tube of finite length $L$, we employ open
boundary conditions for the electrons, $\Pi_{j}(0)=\Pi_{j}(L)=0$, and
$\phi_{j}(L)-\phi_{j}(0) = 0$
\cite{rf:Kane9712,rf:Fabrizio9506,rf:Mattsson9712}.
For the phonon field we have similar boundary conditions, $P(0)=P(L)=0$,
$u(0)=u(L)=0$.
The discreteness of the phonon and electron spectrum in a finite length
SWNT is better visualized by expanding each field operator in Fourier
series as
\begin{eqnarray}
 \phi_{j}(x) & = &
 \sqrt{ \frac{\hbar g_{j}}{L} } \sum_{n \ge 1} \sin(k_{n} x)
 \frac{1}{\sqrt{k_{n}}}
 (b_{j,n}^{\dagger} + b_{j,n} ),
 \label{eq:phi_x} \\
 \Pi_{j}(x) & = &
 i \sqrt{ \frac{\hbar}{g_{j} L} } \sum_{n \ge 1} \sin(k_{n} x)
 \sqrt{ k_{n} }
 (b_{j,n}^{\dagger} - b_{j,n} ). \label{eq:Pi_x}
\end{eqnarray}
Here $b_{j,n}$ ($b_{j,n}^{\dagger}$) are annihilation (creation)
operators of the $j$-sector's bosons, satisfying
$[b_{j,n},b_{j^{'},n^{'}}^{\dagger}]=\delta_{jj^{'}} \delta_{nn^{'}}$,
and $k_{n}=\pi n/L$ is the wave number. Similar expansions hold for
$u(x)$ and $P(x)$:
\begin{eqnarray}
 u(x) & = &
 \sqrt{ \frac{\hbar}{\zeta v_{\rm st} L} } \sum_{n \ge 1} \sin(k_{n} x)
 \frac{1}{\sqrt{k_{n}}} (a_{n}^{\dagger} + a_{n} ), \\
 P(x) & = &
 i \sqrt{ \frac{\hbar \zeta v_{\rm st}}{L} } \sum_{n \ge 1} \sin(k_{n} x)
 \sqrt{ k_{n} } (a_{n}^{\dagger} - a_{n} ),
\end{eqnarray}
where $a_{n}$ ($a_{n}^{\dagger}$) are the annihilation (creation)
operators of the phonon mode obeying the commutation relation
$[a_{n},a_{n^{'}}^{\dagger}]=\delta_{nn^{'}}$.
Using these relations we get
\begin{eqnarray}
 H_{\rm el} & = &
  \sum_{j, n \ge 1}  \varepsilon_{j} n b^{\dagger}_{j,n} b_{j,n}
  +H_{N},
  \label{eq:H_j_boson} \\
 H_{\rm L,ph} & = &
  \sum_{n \ge 1}  \varepsilon_{a} n a^{\dagger}_{n} a_{n},
  \label{eq:H_ph_boson} \\
 H_{\rm el-L,ph} & = &
  \sum_{n \ge 1}  I n (a^{\dagger}_{n} + a_{n})
                            (b^{\dagger}_{c+,n} + b_{c+,n}).
  \label{eq:H_el-ph_boson}
\end{eqnarray}
Here $ \varepsilon_{j}= \varepsilon_{0}/g_{j}$ is the discrete
excitation energy for the $j$-sector,
while $\varepsilon_{a}=\hbar v_{\rm st} \pi/L$ for the phonons. Finally,
$I=c \sqrt{\hbar g/ \pi \zeta v_{\rm st} } \pi/L$ is the electron-phonon
coupling.
The electron-phonon interaction term (\ref{eq:H_el-ph_boson}) reflects
the fact that the phonons couple only to the total charge density via
the deformation potential.
The Hamiltonian $H = H_{\rm el} + H_{\rm L,ph} + H_{\rm el-L,ph}$
is bi-linear in the boson operators characterizing the electronic
and phonon excitations, and as such it can be exactly diagonalized
with a Bogoliubov transformation (see Appendix).
The diagonalized Hamiltonian is,
\begin{eqnarray}
 H & = &
  \sum_{n \ge 1}  E_{\beta} n \beta^{\dagger}_{n} \beta_{n}
  + \sum_{n \ge 1}  E_{\alpha} n \alpha^{\dagger}_{n} \alpha_{n}
  + \sum_{j^{'}, n \ge 1}
     \varepsilon_{0} n b^{\dagger}_{j^{'}\neq c_+,n} b_{j^{'},n}
  + H_{N}, \label{eq:H_tot_dia}
\end{eqnarray}
where $ E_{\beta/\alpha}$ comes from the branch of the charge
density/phonon, defined by,
\begin{eqnarray}
 \!\!\! E_{\beta/\alpha} & = &
  \sqrt{ \frac{ \varepsilon_{a}^{2} +  \varepsilon_{{\rm c}+}^{2}}{2}
  \pm \sqrt{ \left( \frac{ \varepsilon_{c+}^{2}
          -  \varepsilon_{a}^{2}}{2} \right)^{2}
  + 4  I ^{2} \varepsilon_{a}  \varepsilon_{{\rm c}+}  }
  }~.  \nonumber\\
  \label{eq:E}
\end{eqnarray}
Here we assumed the relation $ \varepsilon_{a} < \varepsilon_{{\rm
c}+}$, as it holds for carbon nanotubes.
The summation $j^{'}$ in Eq. (\ref{eq:H_tot_dia}) is for the neutral
sectors $j^{'}={\rm c-},{\rm s+},{\rm s-}$. We notice that for
electron-phonon couplings $I^2 = \varepsilon_{a}\varepsilon_{\rm c+}/4$ the
energy $E_\alpha$ vanishes.  For even larger couplings $E_\alpha$
becomes complex which is an unphysical situation. Thus we require that
$I^2 \le \varepsilon_{a}\varepsilon_{\rm c+}/4$, where the equality sign
defines the Wentzel-Bardeen singularity.
We notice that also the diagonalized energies have a linear dispersion
relation.
Hereafter we use the index $\mu=\alpha, \beta, \nu$ for  the three
kinds of bosonic excitations, and set $E_{\nu}=\varepsilon_{0}$.

Let us now also include the effects of the breathing mode.
A quantitative estimate of these effects  is discussed in
\cite{rf:Martino0306}, where the authors evaluate the effective
retarded electronic action obtained upon integrating out the
phonon stretching and breathing modes. It turns out that
retardation effects can be disregarded for the breathing but not
for the stretching mode. Thus the breathing mode yields a
renormalization of the velocity associated to the charged plasmon
mode as $v_{\rm c+}^{*}=a_{\rm B} v_{\rm c+}^{}$, where
\begin{equation}
a_{\rm B} = \sqrt{1 - \frac{R_{\rm B}}{R} g^{2}}, \label{eq:vstar}
\end{equation}
and $R_{\rm B} \simeq 2.4 \pm 0.9 {\rm \AA}$ \cite{rf:Martino0306}. With
$v_{\rm F}=8 \times 10^{5} {\rm m/sec}$, $v_{\rm st}=2 \times 10^{4}
{\rm m/sec}$, and interaction parameter $g=0.2$ the charge velocity
$v_{\rm c+}^{*}$ becomes {\em comparable} to the sound velocity $v_{\rm
st}$ of the stretching mode if $a_{\rm B} \simeq 5\times 10^{-3}$. This
cannot occur, not even for the smallest SWNT radii.  However, the
renormalized charge velocity might become of the order of the 
sound velocity $v_{\rm st}$ for realistic nanotube radii for strongly
screened interaction (yielding values of $g$ close to one).
The breathing mode also yields a renormalization of the
electron-electron interaction parameter as $g^{*}=v_{\rm F}/v_{\rm
c+}^{*}=g/a_{\rm B}$. This results in the increase of the
electron-phonon coupling in Eq. (\ref{eq:H_el-ph_boson}) with
$(I^{*})^2=I^2/a$.
In the following, we shall then consider breathing-mode renormalized
parameters, with a parameter $a_{\rm B}=0.993$ ($g=0.2$), $0.804$
($g=1$)
as for a realistic $(10,10)$ tube.

We notice that also Umklapp scattering processes should reduce the
charge velocity \cite{rf:Schulz9006}. This effect would be dominant near
half-filling, where a strong coupling with the stretching mode is
expected \cite{rf:Loss9410}. Moreover, a suppression of the
electron-phonon coupling is expected at half-filling
\cite{rf:Fabrizio96,rf:Kleinert97}. This situation has not been
considered in our calculation, since SWNTs are usually away from
half-filling \cite{rf:Egger9806}.

\section{Phonon assisted tunneling}

With the help of Eq. (\ref{eq:H_tot_dia}) we can now evaluate the
density of states $\rho(\epsilon)$. In particular, we shall focus on the
density of states at a nanotube end $\rho_{\rm end}$, yielding direct
information on the differential conductance.
The density of states is calculated by expressing the electron field
operators in terms of the boson operators
\cite{rf:Egger9712,rf:Kane9712,rf:Egger9806}.
After lengthy but standard calculations, see Appendix, we arrive at the
following finite temperature density of states at the end of the
nanotube:
\begin{eqnarray}
 \rho_{\rm end}(\hbar\omega) & = &
  \sum_{s=\pm}
  \sum_{\lambda, \sigma}
  \sum_{l,m,n} C_{\alpha, l} C_{\beta, m} C_{\nu, n}
  \nonumber \\
  &&
 \times \delta \left( \hbar\omega - s (E_{\lambda, \sigma}^{s}
                              + l  E_{\alpha}
                              + m E_{\beta}
                              + n E_{\nu})
                              \right),
  \label{eq:rho_fl}
\end{eqnarray}
where $s$ is the index for the positive and negative energy region,
$\lambda=\pm$, $\sigma=\pm$ are the band and spin indices, the
summations $l, m, n$ run on all integers.
$E_{\lambda, \sigma}^{+} = \sum_{j} c_{j} E_{j} N_{j} + \lambda
\varepsilon_{0} \delta - e V_{\rm G}$ and $E_{\lambda, \sigma}^{-} =
E_{\rm c+} + 3\varepsilon_{0}/4 - E_{\lambda, \sigma}^{+}$, where
$c_{j}=1, \lambda, \sigma, \lambda \sigma$ for $j={\rm c+, c-, s+, s-}$,
respectively.
When the temperature is much lower than the level spacings, $k_{\rm B} T
\ll E_{\mu}$, the coefficients $C_{\mu, p}$ are written as
\begin{eqnarray}
 C_{\mu, p} & = &
  C_{\mu, p}^{T=0}
  + \gamma_{\mu} {\rm e}^{-  E_{\mu}/k_{\rm B} T}
  \left( C_{\mu, p+1}^{T=0} + C_{\mu, p-1}^{T=0}
         - 2 C_{\mu, p}^{T=0} \right), \\
 C_{\mu, p}^{T=0} & = &
  \frac{ \Gamma(\gamma_{\mu} + p) }
       {p! \Gamma(\gamma_{\mu})} C_{\mu, 0}^{T=0}
  \Theta(p).
\end{eqnarray}
Here $\Gamma$ is the Gamma function and $\Theta$ is the step function.
The coefficient $\gamma_{\mu}$ corresponds to the end-tunneling
exponents in the bosonic correlation functions, with
\begin{equation}
\gamma_{\alpha} =  \frac{1}{4g} \frac{E_{\alpha}}{\varepsilon_{\rm
c+}} \sin^{2} \varphi , \qquad  \gamma_{\beta} = \frac{1}{4g}
\frac{E_{\beta}}{\varepsilon_{\rm c+}} \cos^{2} \varphi,
\end{equation}
and $\gamma_{\nu}=3/4$. The phase $\varphi$ is defined by
$\varphi = \arctan(-4I\sqrt{\varepsilon_{a}\varepsilon_{\rm c+}}
/(\varepsilon_{a}^{2}-\varepsilon_{\rm c+}^{2}))/2$. Thus, from
(\ref{eq:rho_fl}), one sees that at low temperatures a signature
of the finite length of the SWNT appears in the form of discrete
peaks in the density of states. More precisely, three peak
families with distinct periods are identified, cf. Fig. 1,
reflecting (i) the discrete excitations of the (renormalized)
charge density with energy spacing $E_{\beta}$, (ii) the
excitations of the neutral sectors with spacing $E_{\nu}$, and
(iii) the phonon-assisted-like excitations with spacing
$E_{\alpha}$.
At zero temperature the $l$-th phonon peak height with respect to
that of the corresponding electron excitation peak is given by
$C_{\alpha,l}^{T=0}/C_{\alpha,0}^{T=0} =
\gamma_{\alpha}(\gamma_{\alpha}+1)\cdots(\gamma_{\alpha}+l-1)/l!$.
For example, the ratio for $l=1$ is simply $\gamma_{\alpha}$. For
large $\gamma_\mu$ it holds the asymptotic relation
 \begin{equation}
 C_{\mu,
p}^{T=0} \simeq
  \frac{ \gamma_{\mu}^p }
       {p!} C_{\mu, 0}^{T=0}
  \Theta(p)\;,
  \label{eq:asymptotic}
\end{equation}
   which resembles the relation among satellite peaks
   due to vibron-assisted tunneling as in \cite{rf:Braig0311}.
For large $p$, on the other hand, each peak series shows power-law
in its height, because the relation $C_{\mu, p}^{T=0} \propto
p^{\gamma_{\mu}-1}$ holds if one uses the asymptotic expansion for
the Gamma function.

\begin{figure}[hbtp]
   \begin{center}
   \includegraphics[width=8cm]{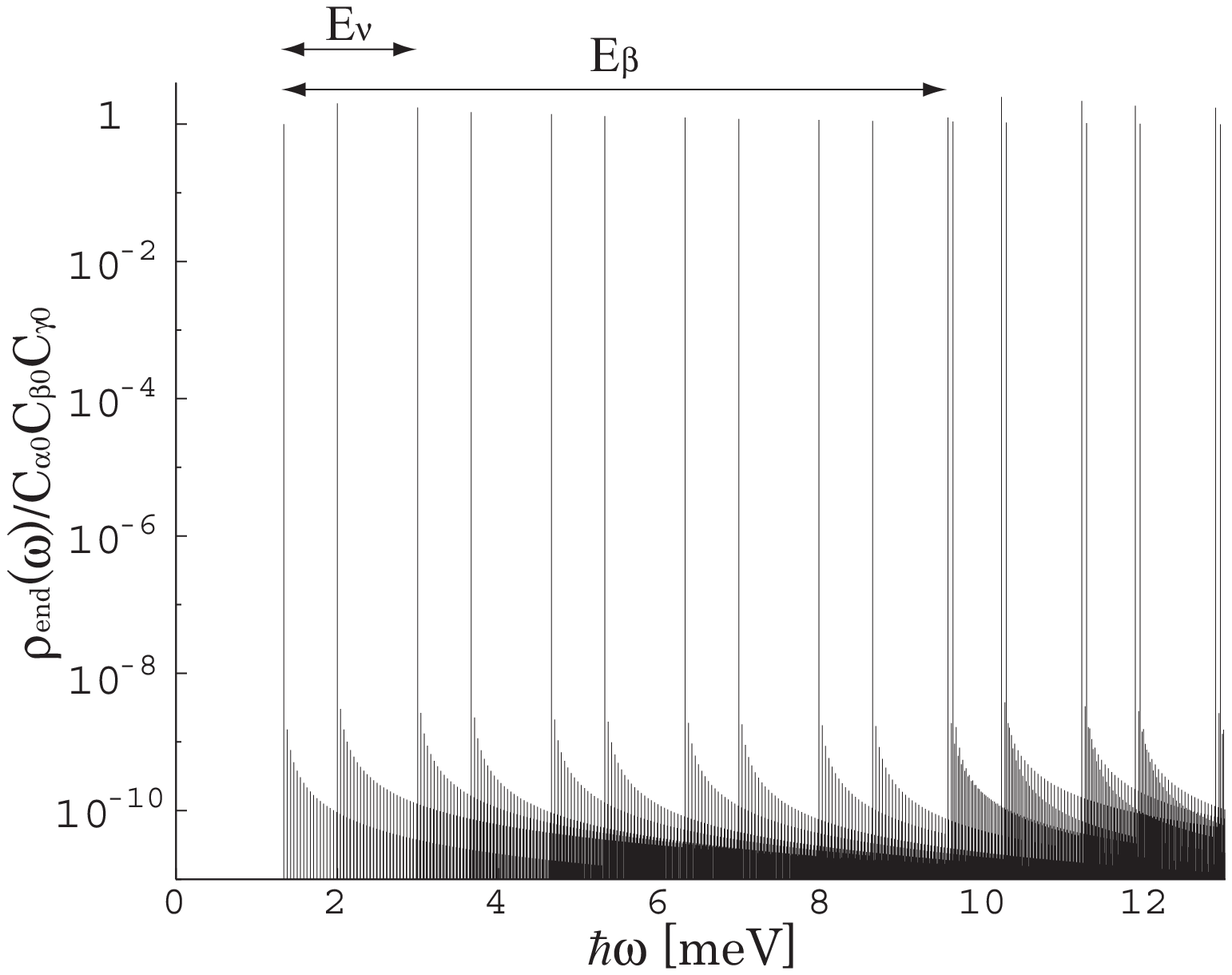}\includegraphics[width=8cm]{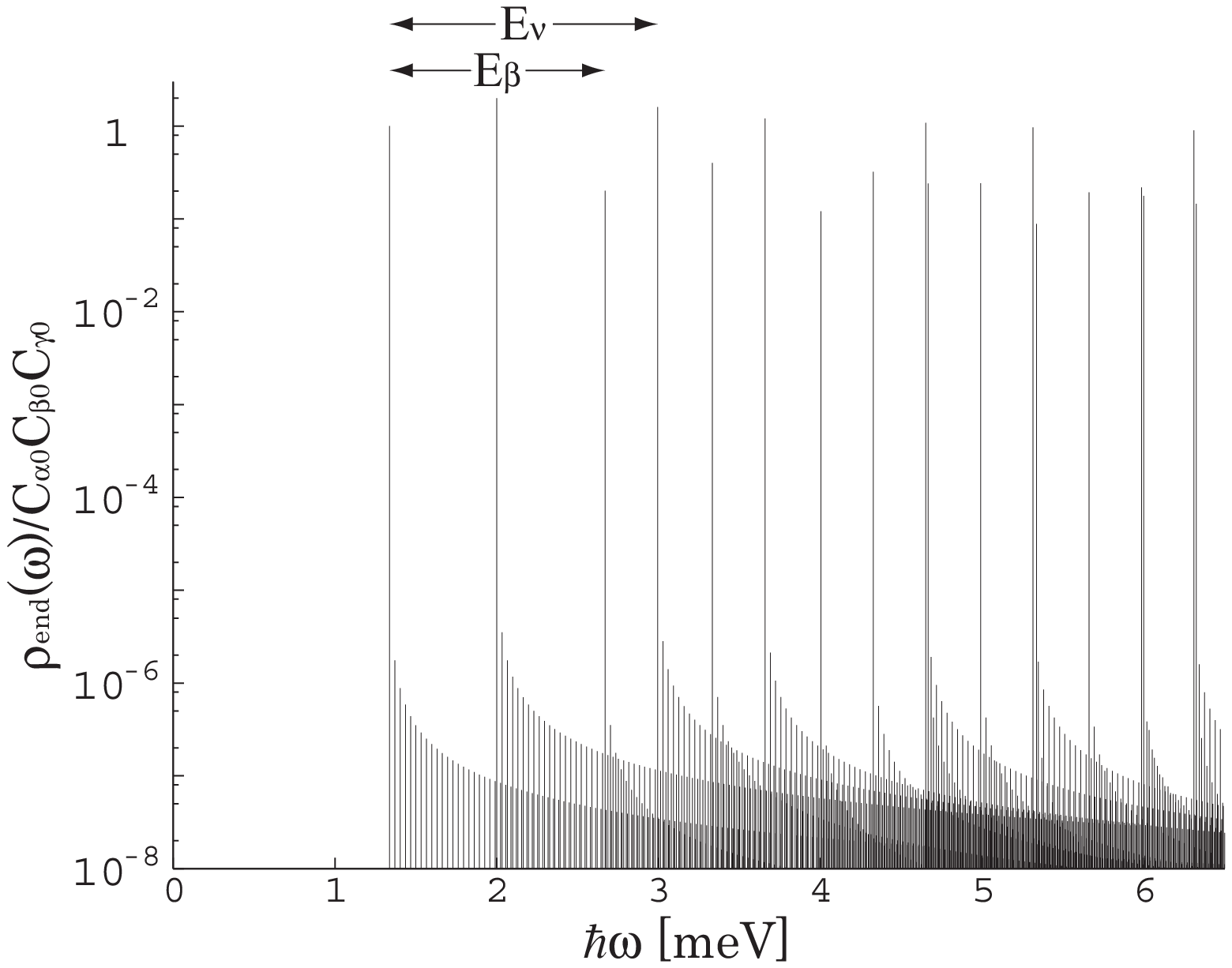}
   \end{center}
   \caption
{End-tunneling density of states of a $(10,10)$ SWNT vs. frequency for
strongly interacting electrons ($g=0.2$, left figure) and
non-interacting electrons ($g=1$, right figure).  Only the positive
energy region is shown.
We use the parameters  $E_{++}^{+}=E_{+-}^{+}=2 {\rm meV}$,
$E_{-+}^{+}=2.99 {\rm meV}$, and $E_{--}^{+}=1.34 {\rm meV}$ for
both figures.  This parameter set corresponds to the condition
$N_{\rm c-}=-1$, $N_{\rm s+}=1$, $N_{\rm s-}=-1$, and finite band
off-set $\delta=0.2$, in which there is an unpaired electron in
the $\lambda=-$ band so that a spin-1/2 appears in the SWNT.
The no peak region in $0 \le \hbar \omega \le 1.34$ is the Coulomb
blockabe region.  The large three peaks at $1.34$, $2$ and $2.99
{\rm meV}$ reflect the different addition energy to each level in
the  SWNT. For  each of these addition energy peaks, three
families of peaks are clearly identified. They are associated to
the plasmon, phonon and "neutral" bosonic excitations of the
suspended SWNT $E_{\beta}$, $E_{\alpha}$ and $E_{\nu}$,
respectively.
Because the relation $E_{-+}^{+}=E_{--}^{+}+E_{\nu}$ holds for the
present electron configuration, some of the excitation peaks
stemming from  $\hbar \omega = E_{--}^{+}$ are degenerate with
some of those stemming from $\hbar \omega = E_{-+}^{+}$.
The remaining parameters for the left figure are
$E_{\alpha}=0.0410{\rm meV}$, $E_{\beta}=8.23{\rm meV}$,
$E_{\nu}=1.66{\rm meV}$, $\gamma_{\alpha}= 1.52 \times 10^{-9}$,
$\gamma_{\beta}= 1.24$, $\gamma_{\nu}= 3/4$, $T=0$.
For the right figure, $E_{\alpha}=0.0327{\rm meV}$, $E_{\beta}=1.33{\rm
meV}$, $E_{\nu}=1.66{\rm meV}$, $\gamma_{\alpha}= 1.76 \times 10^{-6}$,
$\gamma_{\beta}= 0.201$, $\gamma_{\nu}= 3/4$, $T=0$.}
\label{fig:01}
\end{figure}
Here we estimate the amplitudes of the phonon peaks and the excitations'
spacings.
First, we use the following quantities
\cite{rf:Kane9712,rf:Suzuura0205}: $v_{\rm F}=8 \times
10^{5} {\rm m/sec}$, $v_{\rm st}=1.99 \times 10^{4} {\rm m/sec}$,
$c=20 {\rm eV}$, $R=6.79 {\rm \AA}$ (for a $(10,10)$ armchair
SWNT), $\zeta=2 \pi R M$, ($M=3.80 \times 10^{-7} {\rm kg/m^{2}}$),
and $L=1.0 \times 10^{-6} {\rm m}$.
These values give (for $g=0.2$ then $a_{\rm B} = 0.993$) the energies
$E_{\alpha}=0.0410{\rm meV}$, $E_{\beta}=8.23{\rm meV}$,
$E_{\nu}=1.66{\rm meV}$, $\gamma_{\alpha}= 1.52 \times 10^{-9}$, and
$\gamma_{\beta}= 1.24$,
The estimated peak ratio $\gamma_{\alpha}$ is quite small.  However, we
note that the value is sensitive to the parameters.  We may have $g \sim
1$ for the screening caused by the gate electrode.  For $g=1$ ($a_{\rm
B}=0.804$), $E_{\alpha}=0.0327{\rm meV}$, $E_{\beta}=1.33{\rm meV}$,
$E_{\nu}=1.66{\rm meV}$, $\gamma_{\alpha}=1.76 \times 10^{-6}$, and
$\gamma_{\beta}=0.201$.  The sensitivity of $\gamma_{\alpha}$ will be
discussed in the next paragraph.
We also note that there is an uncertainty in the precise value of the
coupling constant $c$. Estimates in \cite{rf:Suzuura0205} yield values
$c \simeq 14-21 {\rm eV}$.

\begin{figure}[hbtp]
   \begin{center}
   \includegraphics[width=8.5cm]{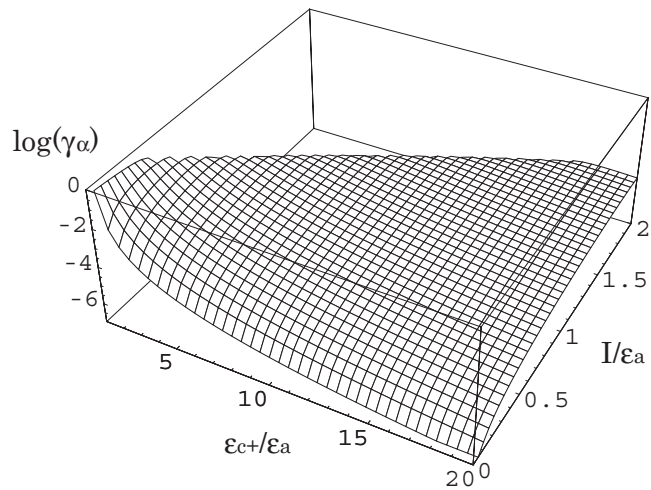}
   \end{center}
   \caption
   {Relative phonon peak height $\gamma_\alpha$ as a function of the
   ratio between charge-plasmon and longitudinal phonons level
   spacing
 $\varepsilon_{\rm c+}/\varepsilon_{a}$, and
 of the electron-phonon coupling $I/\varepsilon_{a}$.
 The vertical axis is in logarithmic scale.
The back side region without mesh is the WB singularity region, where
$\gamma_\alpha$ has an imaginary value.
} \label{fig:02}
\end{figure}

In Fig. \ref{fig:02}, the ratio of the first phonon peak height
$\gamma_{\alpha}$ as a function of the charge plasmon energy
spacing and of the electron-phonon coupling is plotted.
For very strong couplings, corresponding to the condition $I^{2} >
\varepsilon_{a} \varepsilon_{\rm c+}/4$, the system becomes unstable, as
the Wenztel-Bardeen (WB) singularity is approached.
In general the ratio is very sensitive to the parameters.  The peak
height becomes larger for small charge velocities, especially close to
the WB singularity.
For the small coupling case, the system survives the collapse for a wide
range of parameters while the charge velocity is increased, until
finally the approach of the singularity is expected.

\section{Conclusion}

From our semi-quantitative discussion, it follows that the phonon peaks
could be observed in experiments.
In recent experiments \cite{rf:Sapmaz.pre} on suspended metallic carbon
nanotubes, many excitation peaks have been observed in the differential
conductance.  The peaks appear outside of the Coulomb diamonds, and are
almost equidistantly spaced.
The period of the peaks is about $0.53 {\rm meV}$ for a $L=140 {\rm nm}$
length nanotube, $0.69 {\rm meV}$ for a $L=420 {\rm nm}$, and $0.14 {\rm
meV}$ for a $L=1200 {\rm nm}$, respectively.  The theoretical estimate
of the period for the phonon assisted peaks is $0.77$ to $0.09 {\rm
meV}$, while that for the charge density is $47.5$ to $5.5 {\rm meV}$
(if $g \sim 0.2$), and $9.5$ to $1.1 {\rm meV}$ for the neutral sectors.
The observed peaks might be the phonon-assisted peaks discussed in this
paper.
The relative peak height $\gamma_\alpha$ reflects the strength of
the electron-phonon coupling. From the experimental data in Ref.
\cite{rf:Sapmaz.pre} a rough estimate $\gamma_{\alpha} \sim 0.3$
to $0.8$ can be extracted. These values are much larger than those
we estimated in the previous section for strong Coulomb
interaction ($g \sim 0.2$). Moreover, the experiments seem to well
agree with the asymptotic formula (\ref{eq:asymptotic}). This
could be an indication of a strong renormalization of the
electronic parameters expected when the system is close to the WB
singularity. Thus, while a qualitative agreement with the
experiments is found, an understanding of the lack of a
quantitative agreement clearly calls for further theoretical
investigations.
We mention that a relation similar to (\ref{eq:asymptotic}) has
also been predicted for electronic degrees of freedom dressed by a
Frank-Condon factor originating from a generic vibron-electron
coupling. However, also the theoretical expectation for the values
of the coupling in the model in \cite{rf:Braig0311}  seems to be
much smaller than the one extracted from experiments
\cite{rf:Sapmaz.pre}.

For temperatures larger than the level spacing finite size effects are
washed out, and we can adapt our calculation to the case of a SWNT of
infinite length.  The density of states for this case becomes:
\begin{eqnarray}
\rho_{\rm end} (\hbar\omega) & \sim & |\omega |^{\frac{1}{4}
\left( \frac{1}{g'}+3 \right)  -1},
\end{eqnarray}
where $g'$ is the interaction parameter renormalized by the
electron-phonon interaction,
$(g')^{-1}=4(\gamma_{\alpha}+\gamma_{\beta})$.
As expected, the charge, neutral and phonon peaks are washed out and the
density of states shows a simple power-law behavior with exponent
modified by the electron-phonon interaction.  The renormalized parameter
$g'$ is always larger than $g$, showing that the phonons contribute to
an effective {\em attractive} electron-electron interaction.

When a gate voltage is applied to the nanotube, the possibility to
investigate Coulomb blockade phenomena and Kondo-correlation
effects arises.
In the presence of a gate potential the coupling between the
electrons and the transverse bending mode should be included.  One
can control the coupling strength by changing the gate potential.
Because the bending phonon mode couples to many  of the charge
density modes, and it has less energy for the vibration, an
instability might be easily induced.
The effects of the gate potential will be discussed elsewhere.

In conclusion, we have discussed transport properties of a suspended
metallic SWNT.  The low energy excitations caused by the stretching
phonons are expected to appear as small periodic peaks beside the peaks
for the charged and neutral electronic-like modes in the differential
conductance.  Each series of peaks exhibit a different power-law.  When
the system is close to the WB singularity, the phonon peaks are large
enough in height to be observed. The observation of the phonon peaks
would thus indicate that the singularity is approached.

\section*{Acknowledgments}

We thank S. Sapmaz, Y. M. Blanter, L. Mayrhofer, R. Saito for valuable
discussions.
This work was supported by the Fundamental Research on Matter in
the Netherlands; by the Grant-in-Aid No. 16740166 from the
Ministry of Education, Culture, Sport, Science and Technology, of
Japan; by the Deutsche Forschungsgemeinschaft (GRK 638).

\section*{Appendix: Density of states of suspended SWNTs}

Here we show the detailed calculation of the density of states at
the end of a finite length nanotube.  Let us consider a high
enough  tunneling barrier between  electrodes and the nanotube. At
low temperature, the differential conductance is proportional to
the density of states at the end of the nanotube
\begin{equation}
dI/dV_{\rm bias} \propto \rho_{\rm end}(\hbar\omega = e V_{\rm
bias} ) = - {\rm Im} G_{\rm end}(\hbar\omega= e V_{\rm bias}) /
\pi \;,
\end{equation}
where $G_{\rm end}(\hbar \omega)\equiv G(x,x';\hbar \omega)|_{x=x'
\rightarrow \xi}$ is the Fourier transform of the retarded Green's
function $G(x,x';t) = \frac{1}{i \hbar} \Theta(t) \langle \Psi(x,t)
\Psi^{\dagger}(x',0) + \Psi^{\dagger}(x',0) \Psi(x,t) \rangle$, and
$\xi$ is a length which is of the order of the inverse of the Fermi wave
length.
The electron field operator is written in terms of boson operators using
the bosonization identity \cite{rf:comment.3D}
\begin{eqnarray}
\Psi_{\eta}(x,t) & = & \frac{1}{\sqrt{2 \pi \epsilon}}
                       U_{\eta}
                       {\rm e}^{i \kappa_{\eta} x}
                       {\rm e}^{- \frac{i}{2}
                                \sum_{j} c_{j} \theta_{j, r}(x,t)},
\end{eqnarray}
and $\Psi(x,t)=\sum_{\eta} \Psi_{\eta}(x,t)$. Here $\eta=(r,
\lambda, \sigma)$ is the parameter set of the left and right going
wave ($r=-, +$), the band index of the nanotube ($\lambda=\pm$),
and the spin index ($\sigma=\pm$).
$\epsilon \rightarrow 0$ is the cut-off length, $U_{\eta}$ is the Klein
factor, $\kappa_{\eta} = r \lambda k_{\rm F} + r ( N_{\eta} + \lambda
\delta )\pi/L $ and $N_{\eta}$ is the number of electrons in the
$\eta$-branch.
$k_{\rm F}=2 \pi/3 a$ is the Fermi wave length, $a$ is the honeycomb
lattice constant, and we consider the case $k_{\rm F} \gg |N_{\eta}|
\pi/L$ in this paper to justify our low-energy theory.
The field $\theta_{j, r}$ is written as,
\begin{eqnarray}
\theta_{j, r}(x,t) & = & \sqrt{ \frac{\pi}{\hbar} }
                        \left( - r \phi_{j} (x,t)
                               + \int^{x} dx' \Pi_{j} (x') \right).
\label{eq:theta_x}
\end{eqnarray}
We note that the left and right going waves are not independent
anymore in the case of open boundary condition; the relation
$\Psi_{r, \lambda, \sigma}(x) = - \Psi_{-r, \lambda, \sigma}(-x)$
holds.

Let us consider the correlation function in the Green's function
$\langle \Psi(x,t) \Psi^{\dagger}(x',0) \rangle$.  This can be separated
into the boson and the fermion parts;
\begin{eqnarray}
\langle \Psi(x,t) \Psi^{\dagger}(x',0) \rangle & = &
\sum_{\eta} \frac{1}{2 \pi \epsilon}
\langle U_{\eta}(t) U_{\eta}^{\dagger}(0) \rangle
\nonumber \\
&&
\times
\left\{
{\rm e}^{i \kappa_{\eta} (x-x')}
\Pi_{j} \langle {\rm e}^{-i \frac{1}{2} c_{j} \theta_{j, r}(x,t)}
                {\rm e}^{ i \frac{1}{2} c_{j} \theta_{j, r}(x',0)} \rangle \right.
\nonumber \\
&&
\left.
-
{\rm e}^{i \kappa_{\eta} (x+x')}
\Pi_{j} \langle {\rm e}^{-i \frac{1}{2} c_{j} \theta_{j, r}(x,t)}
                {\rm e}^{ i \frac{1}{2} c_{j} \theta_{j, r}(-x',0)} \rangle
\right\}.
\end{eqnarray}
Since the Klein factor only changes the number of electrons in
$\eta$-branch, the fermion part of the correlation function can be
calculated as,
$\langle U_{\eta}(t) U_{\eta}^{\dagger}(0) \rangle = {\rm exp}(- i
\frac{E_{N_{\eta} + 1} - E_{N_{\eta}}}{\hbar} t )$.
Here $E_{N_{\eta} + 1} - E_{N_{\eta}} = E_{\lambda, \sigma}^{+}$
being addition energies of the $\eta$-branch.  (Because of the
open boundary condition, we have no $r$ dependence in the addition
energies).
On the other hand, using the relation for bosonic operators,
$\langle {\rm e}^{- i A} {\rm e}^{i B} \rangle =
{\rm e}^{-\frac{1}{2} \langle A^{2} \rangle
         -\frac{1}{2} \langle B^{2} \rangle
         + \langle AB \rangle }$,
the bosonic part of the correlation function can be written as
\begin{eqnarray}
&& \langle {\rm e}^{- \frac{i}{2} c_j \theta_{j, r}(x, t)}
           {\rm e}^{  \frac{i}{2} c_j \theta_{j, r}(x',0)} \rangle
   \nonumber \\
&&  = {\rm e}^{ \{ - \frac{1}{2} \langle \theta_{j, r}^{2}(x,t)
\rangle - \frac{1}{2} \langle \theta_{j, r}^{2}(x',0) \rangle +
\langle \theta_{j, r}(x,t) \theta_{j, r}(x',0) \rangle \}/4 },
\end{eqnarray}
which doesn't depend on $\lambda$, $\sigma$.  Therefore the calculation
can be reduced to that for the bosonic correlation function, $\langle
\theta_{j, r}(x,t) \theta_{j, r}(x',0) \rangle$.
Now we discuss the correlation function of the charge sector.
Similar calculations have been done for the neutral sectors.
Substituting the Eqs. (\ref{eq:phi_x}), (\ref{eq:Pi_x}) into
(\ref{eq:theta_x}), and using the Bogoliubov transformation
\begin{eqnarray}
\left(
\begin{array}{c}
\beta_{n}^{\dagger} \\
\beta_{n} \\
\alpha_{n}^{\dagger} \\
\alpha_{n} \\
\end{array}
\right)
& = &
\hat{A_{n}} \hat{B_{n}} \hat{C_{n}}
\left(
\begin{array}{c}
b_{{\rm c+}, n}^{\dagger} \\
b_{{\rm c+}, n} \\
a_{n}^{\dagger} \\
a_{n} \\
\end{array}
\right),
\end{eqnarray}
where
\begin{eqnarray}
\hat{A_{n}} & = &
\left(
\begin{array}{cccc}
\frac{1}{\sqrt{ E_{\beta} n }} & -i \sqrt{ E_{\beta} n} & 0 & 0 \\
\frac{1}{\sqrt{ E_{\beta} n }} &  i \sqrt{ E_{\beta} n} & 0 & 0 \\
0 & 0 & \sqrt{ E_{\alpha} n} & -\frac{i}{\sqrt{ E_{\alpha} n }} \\
0 & 0 & \sqrt{ E_{\alpha} n} &  \frac{i}{\sqrt{ E_{\alpha} n }} \\
\end{array}
\right), \\
\hat{B_{n}} & = &
\left(
\begin{array}{cccc}
\cos \varphi & 0 & 0 & -\sin \varphi \\
0 &  \cos \varphi & \sin \varphi & 0 \\
0 & -\sin \varphi & \cos \varphi & 0 \\
\sin \varphi & 0 & 0 &  \cos \varphi \\
\end{array}
\right), \\
\hat{C_{n}} & = &
\left(
\begin{array}{cccc}
\frac{i}{2} \sqrt{ \varepsilon_{\rm c+} n } & -\frac{i}{2} \sqrt{
 \varepsilon_{\rm c+} n } & 0 & 0 \\
-\frac{1}{2 \sqrt{ \varepsilon_{\rm c+} n } } & -\frac{1}{2 \sqrt{
 \varepsilon_{\rm c+} n } } & 0 & 0 \\
0 & 0 & \frac{1}{2 \sqrt{ \varepsilon_{a} n } } & \frac{1}{2 \sqrt{
 \varepsilon_{a} n } } \\
0 & 0 & \frac{i}{2} \sqrt{ \varepsilon_{a} n }& -\frac{i}{2} \sqrt{
 \varepsilon_{a} n } \\
\end{array}
\right), \\
\end{eqnarray}
at the end of nanotube $x=x' \rightarrow \xi$, the correlation function
satisfies the relation
\begin{eqnarray}
&&
\langle \theta_{{\rm c+}, r}(x,t) \theta_{{\rm c+}, r}(x',0) \rangle
\nonumber \\
& = & \frac{1}{g_{j}} \frac{\pi}{L} \sum_{n \ge 1} \frac{1}{k_{n}}
\left\{ \frac{E_{\beta}}{\varepsilon_{\rm c+}} \cos^{2} \varphi
\langle \beta_{n}(t) \beta_{n}^{\dagger}(0)
+  \beta_{n}^{\dagger}(t) \beta_{n}(0) \rangle \right. \nonumber \\
&& +
\left.
\frac{E_{\alpha}}{\varepsilon_{\rm c+}}
\sin^{2} \varphi
 \langle \alpha_{n}(t) \alpha_{n}^{\dagger}(0)
+  \alpha_{n}^{\dagger}(t) \alpha_{n}(0) \rangle \right\},
\label{eq:boson_cor}
\end{eqnarray}
which doesn't depend on $r$.
The lowest order of the function with respect to $\xi$ is $O(\xi^0)$,
therefore we calculate the bosonic correlation function at $x=x'=0$
hereafter.
Using the relation
\begin{eqnarray}
\langle \beta_{n}(t) \beta_{n}^{\dagger}(0)
+ \beta_{n}^{\dagger}(t) \beta_{n}(0) \rangle
& = & \cos(E_{\beta} n t)
\cosh \left( \frac{E_{\beta} n}{2 k_{\rm B} T} \right)
- i \sin(E_{\beta} n t),
\end{eqnarray}
and multiplying by ${\rm e}^{- \hbar v k_{n} /E_{\rm cut}}$ each term of
the summation over $n$ in Eq. (\ref{eq:boson_cor}) to avoid divergences
in the calculation, where $E_{\rm cut} = \hbar v \epsilon^{-1}
(\rightarrow \infty)$ is a cut-off energy and $v = E_{\alpha / \beta} L
/ \pi \hbar$,
then the exponent of the Green's function
is written as
\begin{eqnarray}
&&
\langle \theta_{j, r}^{2}(0,t) \rangle
+ \langle \theta_{j, r}(0,t) \theta_{j, r}(0,0) \rangle \nonumber \\
& = & - \frac{1}{g_{\rm c+}} \frac{ E_{\beta} }{ \varepsilon_{\rm
c+} } \cos^{2} \varphi \left\{ \log \frac{ 1 - {\rm
e}^{-E_{\beta}/E_{\rm cut}}
                {\rm e}^{-i E_{\beta} t / \hbar} }
          { {\rm e}^{-E_{\beta}/E_{\rm cut}} }
\right.
\nonumber \\
&& \left. + \sum_{m} \log \frac{ \cosh ( \frac{E_{\beta}}{E_{\rm
cut}}
                    + \frac{m E_{\beta}}{k_{\rm B} T} )
          - \cos ( \frac{E_{\beta}}{\hbar} t ) }
          { \cosh ( \frac{E_{\beta}}{E_{\rm cut}}
                    + \frac{m E_{\beta}}{k_{\rm B} T} ) -1 }
\right\} \nonumber \\
&& +
\{
\beta \rightarrow \alpha,
\cos \varphi \rightarrow \sin \varphi
\},
\end{eqnarray}
where we use the identities
\begin{eqnarray}
\cosh x & = & 1 + 2 \sum_{n=1}^{\infty} {\rm e}^{-2 n x}, \\
\sum_{n=1}^{\infty} \frac{ {\rm e}^{- n x} }{n} & = & - \log ( 1 - {\rm e}^{-x} ).
\end{eqnarray}
Then the boson part of the correlation function is calculated as
\begin{eqnarray}
\langle {\rm e}^{- \frac{i}{2} \theta_{j, r}(0,t)}
        {\rm e}^{\frac{i}{2} \theta_{j, r}(0,0)} \rangle
& = & D_{\alpha}(t) D_{\beta}(t),
\end{eqnarray}
where, up to orders $(\exp \{-E_\beta/k_BT\})$,
\begin{eqnarray}
D_{\beta}(t) & = &
\left(
\frac{ 1 - {\rm e}^{-E_{\beta}/E_{\rm cut}}
                {\rm e}^{-i E_{\beta} t / \hbar} }
          { {\rm e}^{-E_{\beta}/E_{\rm cut}} }
\right)^{-\gamma_\beta}
\nonumber \\
&&
\times
\Pi_{m}
\left(
\frac{ \cosh ( \frac{E_{\beta}}{E_{\rm cut}}
                    + \frac{m E_{\beta}}{k_{\rm B} T} )
          - \cos ( \frac{E_{\beta}}{\hbar} t ) }
          { \cosh ( \frac{E_{\beta}}{E_{\rm cut}}
                    + \frac{m E_{\beta}}{k_{\rm B} T} ) -1 }
\right)^{-\gamma_\beta}
\\
& = &
\sum_{p=-\infty}^{\infty} D_{\beta, p}
{\rm e}^{-i p \frac{E_\beta}{\hbar} t },
\end{eqnarray}
where
\begin{eqnarray}
 D_{\beta, p} & = &
  D_{\beta, p}^{T=0}
  + \gamma_{\beta} {\rm e}^{-  E_{\beta}/k_{\rm B} T}
  \left( D_{\beta, p+1}^{T=0} + D_{\beta, p-1}^{T=0}
         - 2 D_{\beta, p}^{T=0} \right), \\
 D_{\beta, p}^{T=0} & = &
  \frac{( 1 - {\rm e}^{-E_{\beta}/E_{\rm cut}} )^{\gamma_\beta}}
       { E_{\beta} / \hbar}
  {\rm e}^{-p E_{\beta}/E_{\rm cut}}
  \frac{ \Gamma(\gamma_{\beta} + p) }
       {p! \Gamma(\gamma_{\beta})}
  \Theta(p),
\end{eqnarray}
and $\beta \rightarrow \alpha$ for $D_{\alpha}(t)$.
After a similar calculation for the neutral sectors and for
$\langle \Psi^{\dagger}(x',0) \Psi(x,t) \rangle$, we finally get
the Green's function
\begin{eqnarray}
G_{\rm end}(\hbar \omega) & = &
\frac{2 \sin^{2} (k_{\rm F} \xi) }{\pi \epsilon}
\sum_{s=\pm}
\sum_{\lambda, \sigma}
\sum_{l,m,n} D_{\alpha, l} D_{\beta, m} D_{\nu, n}
\nonumber \\
&& \times
      \frac{1}{\hbar \omega - s (E_{\lambda, \sigma}^{s}
                   + l E_{\alpha}
                   + m E_{\beta}
                   + n E_{\nu} ) + i 0_{+} }.
\end{eqnarray}
After re-definition of the coefficients, e.g., $C_{\mu, p} = D_{\mu, p}
(2 \sin^{2}(k_{\rm F} \xi) /\pi \epsilon )^{1/3}$, we get the density of
states (\ref{eq:rho_fl}).


\begin{thebibliography}{99}
\bibitem{rf:comment.3D} In this derivation we neglect a slowly
oscillating prefactor which comes by considering the three-dimensional
nature of the nanotube wavefunctions.  See e.g. Ref. \cite{rf:Egger9712}.
\end{thebibliography}
\end{document}